\documentclass[usenatbib]{mnras}

\usepackage{newtxtext,newtxmath}

\usepackage[T1]{fontenc}

\DeclareRobustCommand{\VAN}[3]{#2}
\let\VANthebibliography\thebibliography
\def\thebibliography{\DeclareRobustCommand{\VAN}[3]{##3}\VANthebibliography}

\usepackage{graphicx}	% Including figure files
\usepackage{amsmath}	% Advanced maths commands
\usepackage{amssymb}	% Extra maths symbols
\usepackage{bbm}

\title[wdtools]{Computational Tools for the Spectroscopic Analysis of White Dwarfs}

\author[Chandra, Hwang, Zakamska \& Budav\'ari]{
Vedant Chandra,$^{1}$\thanks{E-mail: \href{mailto:vchandra@jhu.edu}{vchandra@jhu.edu}}
Hsiang-Chih Hwang,$^{1}$
Nadia L. Zakamska,$^{1}$
and Tam\'as Budav\'ari$^{1,2}$
\\
$^{1}$Department of Physics \& Astronomy, Johns Hopkins University, 3400 N Charles St, Baltimore MD 21218\\
$^{2}$Department of Applied Mathematics \& Statistics, Johns Hopkins University, 3400 N Charles St, Baltimore MD 21218
}

\date{Accepted 2020 July 21. Received 2020 July 20; in original form 2020 June 17}

\pubyear{2020}

\begin{document}
\label{firstpage}
\pagerange{\pageref{firstpage}--\pageref{lastpage}}
\maketitle
\begin{abstract}
    The spectroscopic features of white dwarfs are formed in the thin upper layer of their stellar photosphere. These features carry information about the white dwarf's surface temperature, surface gravity, and chemical composition (hereafter `labels'). Existing methods to determine these labels rely on complex ab-initio theoretical models which are not always publicly available. Here we present two techniques to determine atmospheric labels from white dwarf spectra: a generative fitting pipeline that interpolates theoretical spectra with artificial neural networks, and a random forest regression model using parameters derived from absorption line features. We test and compare our methods using a large catalog of white dwarfs from the Sloan Digital Sky Survey (SDSS), achieving the same accuracy and negligible bias compared to previous studies. We package our techniques into an open-source Python module `{\sc wdtools}' that provides a computationally inexpensive way to determine stellar labels from white dwarf spectra observed from any facility. We will actively develop and update our tool as more theoretical models become publicly available. We discuss applications of our tool in its present form to identify interesting outlier white dwarf systems including those with magnetic fields, helium-rich atmospheres, and double-degenerate binaries.
\end{abstract}
	
\begin{keywords}
    white dwarfs -- techniques: spectroscopic -- software: data analysis
\end{keywords}
	
\section{Introduction}
	
Deriving precision stellar parameters for white dwarfs, primarily their effective temperature $T_{\text{eff}}$ and gravity $\log g$, has been subject of intense theoretical effort by multiple groups \citep{Hubeny1988, Fontaine2001, Koester2008, Tremblay2009, Tremblay2013}. These labels are of immense interest for several areas of astrophysics. For example, a lower bound on the age of the Universe can be derived by modeling cooling tracks of the oldest white dwarfs in globular clusters \citep{Hansen2002}. Another application is the search for near- and super-Chandrasekhar close binaries \citep{Brown2013} and for massive white-dwarf merger products which can constrain the double-degenerate scenario of type Ia progenitors \citep{Maoz2013}. 
	
\begin{figure}
    \centering
    \includegraphics[width=\columnwidth]{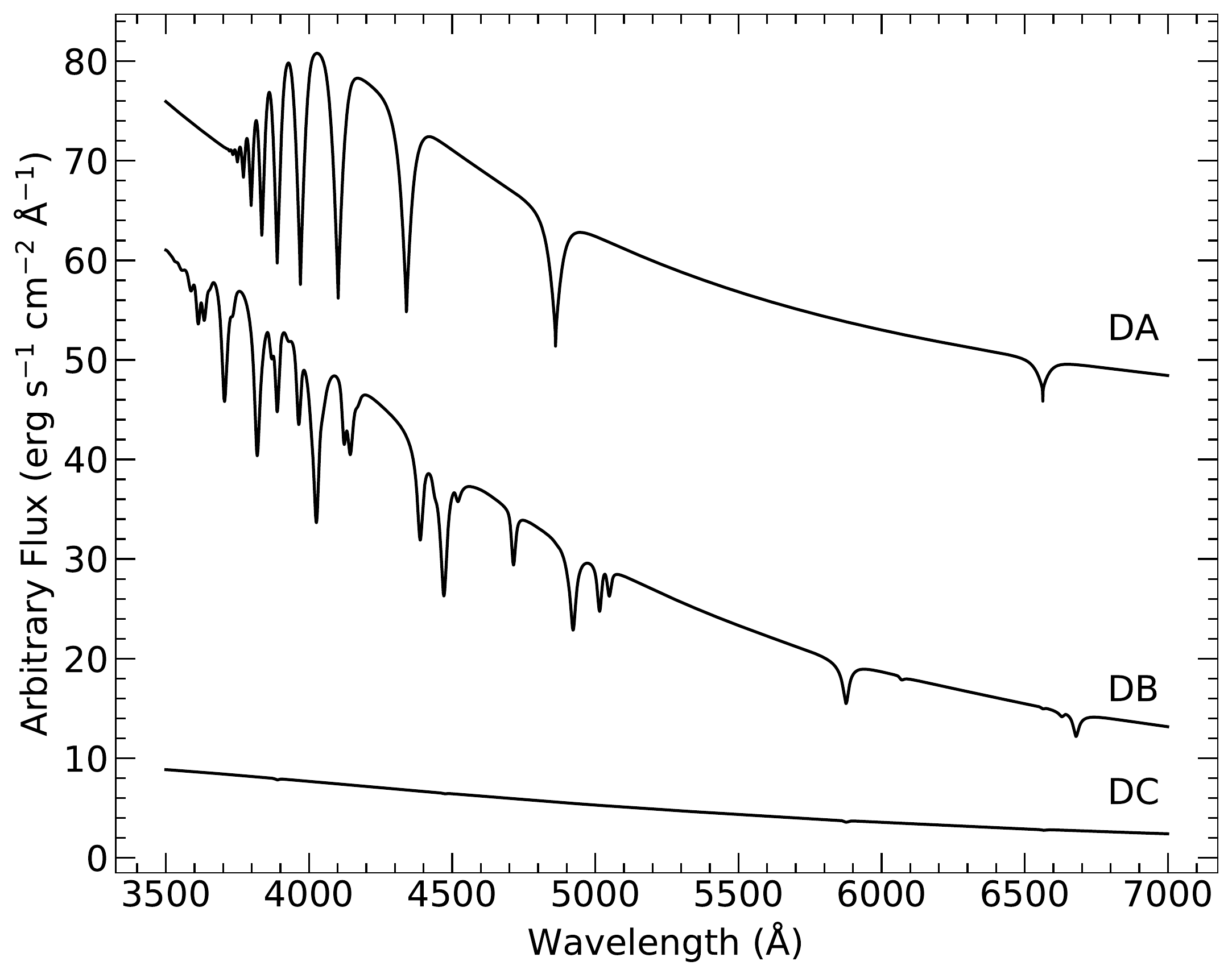}
    \caption{Three un-normalized model spectra from \citet{Koester2010} with DA (hydrogen-rich), DB (helium-rich) and DC (continuum) spectral classifications respectively, vertically displaced by a constant for clarity. The DA spectra have their own set of synthetic models, whilst the DB and DC spectra are taken from a common grid of models. From top to bottom, these model spectra have ($T_{\text{eff}}$, $\log{g}$) labels $(16500, 6.75)$, $(17000, 7)$, and $(9250, 7)$ respectively.}
    \label{fig:specclass}
\end{figure}

For hydrogen-rich (DA) white dwarfs, some of the most accurate determinations of white dwarf temperatures and surface gravities are made by fitting theoretical spectra to absorption lines in the Balmer series of the optical range (Figure \ref{fig:specclass}). The widths and equivalent widths of the profiles are sensitive both to temperature and gravity. Comparisons between spectroscopic and photometric methods \citep{Tremblay2019} indicate that for a wide range of temperatures, the spectroscopic temperature measurements are in agreement to within 2\% and the surface gravity measurements are in agreement within $0.2$ dex. Comparing measurements made by different groups \citep{Tremblay2010} also yields broadly consistent results. 

However, the theoretical models remain uncertain for certain situations, including low temperatures \citep{Tremblay2013, Blouin2020}, high temperatures \citep{Levenhagen2017}, and high masses \citep{Camisassa2019}. Masses are known to be consistently over-estimated at the low-temperature end \citep{Tremblay2010} and there is continued work on improving models in this regime \citep{Tremblay2013}. At these low temperatures, $\log{g}$ values disagree with photometric predictions from \textit{Gaia} \citep{Tremblay2019}. An additional hurdle is that almost all the codes used to model white dwarf atmospheres are kept under restricted access by their respective groups. Rarely, limited grids of pre-computed synthetic spectra are publicly available. Two notable exceptions are the freely-available {\sc tlusty} \citep{tlusty} and {\sc tmap} \citep{tmap} codes. However, uncertainties in atmospheric convection at lower temperatures restrict their usage to hot white dwarfs with $T_{\rm eff} \gtrsim 20,000$ K.

For main-sequence stars, machine-learning based techniques have become a promising solution to predict elemental abundances. For example, the {\sc payne} \citep{Ting2018} uses a neural-network to reproduce spectral models at high spectral resolution, and applies them to lower resolution spectra across a wider range of metal abundances. The {\sc cannon} \citep{Ness2015} uses a data-driven approach to assign astrophysical labels based on stellar features. {\sc astroNN} \citep{Leung2018} uses convolutional neural networks directly on spectral pixels to predict stellar parameters and abundances. These computational approaches are now widely used and have become a powerful technique in astrophysical research \citep{Nataf2019, Xiang2019}. 

White dwarfs are a promising albeit as-yet unexplored population for data-driven techniques. In particular, ~35,000 white dwarf spectra have been measured by the Sloan Digital Sky Survey \citep{Kepler2019}, and Gaia DR2 revealed a large sample of ~250,000 white dwarfs \citep{Fusillo2019}. Recent studies have used follow-up spectroscopy to characterize over a thousand white dwarfs within 100 pc of the Sun \citep{Kilic2020, Tremblay2020, Mccleery2020}. Future large-scale spectroscopic surveys like the Dark Energy Spectroscopic Instrument (DESI; \citealt{DESI2016}) and SDSS-V \citep{Kollmeier2017} will greatly increase the population of white dwarf spectra.

Therefore, there is an urgent need to develop tools to recover the astrophysical parameters of white dwarfs from spectroscopic data and to identify interesting white dwarfs that are inconsistent with theoretical models. \cite{Narayan2019} previously released their fitting routine for DA white dwarfs that uses the high-temperature synthetic spectra from \cite{Bohlin2020} to establish faint WD spectro-photometric standards. Their method includes a sophisticated treatment of extinction, which serves their goal of subpercent photometric accuracy. Our work uses a markedly different methodology for model interpolation and inference, and aims to be more user-friendly and computationally efficient for application to large-scale spectroscopic surveys. 

In this paper, we present an open-source software package in Python called {\sc wdtools} that is meant to be flexible across a wide variety of use-cases, from quick predictions on large datasets to detailed statistical inference for single targets. {\sc wdtools} is hosted online (Zenodo: \citealt{wdtools}) with bug-reporting capabilities from GitHub \footnote{\url{https://github.com/vedantchandra/wdtools} \label{gh}}, along with examples and full code documentation \footnote{\url{https://wdtools.readthedocs.io}}.

Although the model evaluation in this paper is focused on SDSS spectra with known labels, we also show that {\sc wdtools} can be used on spectra obtained from any observatory. The model spectra and SDSS spectra are computed with vacuum wavelengths, and we use that convention in this work. Our tool includes the ability to switch to the air wavelength convention as well. We run all our tests on a MacOS laptop computer with a 3.1 GHz quad-core processor,  and provide computation time estimates for this configuration.

We define our test sample of SDSS white dwarf spectra in Section \ref{Data}. We describe our generative fitting pipeline in Section \ref{synthetic} and our random forest regression model in Section \ref{rf}. We test the accuracy of our tools on a variety of spectra in Section \ref{results} and present applications of our tools to find interesting spectra. We summarize our results and discuss the strengths and limitations of {\sc wdtools} in Section \ref{discussion}. 

\section{Data and Feature Extraction}\label{Data}

\subsection{Data Selection}

We use white dwarf spectra from the Sloan Digital Sky Survey (SDSS) to develop and test our methods. We use stellar parameters derived by \cite{Tremblay2019} as `ground truth' labels as a point of comparison. These were derived using traditional $\chi^2$ fitting with theoretical models to determine temperature and surface gravity, and likely represent the best labels for these stars so far. Alternatives include the catalog of \cite{Fusillo2019}, but we elect to forego those labels since they are derived from \textit{Gaia} photometric observations rather than spectroscopy. Since there is an inherent systematic scatter between photometric and spectroscopic labels \citep{Tremblay2019}, we elect to compare our spectroscopic results to labels derived from spectroscopy as well.

For this study we focus on stars classified by \cite{Tremblay2019} as DA (hydrogen-rich) white dwarfs, which make up to 70 percent of the SDSS sample. We also extend our method of fitting model atmospheres (Section \ref{synthetic}) to DB (helium-rich) white dwarfs which constitute a further 10 percent of the SDSS sample. For the development of our machine learning tools (Section \ref{rf}), we exclude those stars with deviant spectral features like magnetism (H), metallic lines (Z), and pure-continuum spectra (DC). We discuss the possibility of using our fitting tool to identify such outlier spectra in Section \ref{outliers}.

\subsection{Continuum Normalization} \label{contcorrection}

 Due to the lack of absorption lines other than the hydrogen Balmer series, it is relatively straightforward to identify the continuum of a DA white dwarf spectrum. However, to prevent introducing any systematic errors into spectral fitting models, a consistent continuum-normalization technique is vital. We build upon the continuum-normalization technique of \cite{Bergeron1995} and subsequent works, which individually normalize the continuum around each Balmer absorption line with summed Gaussian profiles. 
 
   \begin{table*}
	\centering
	\noindent\makebox[\textwidth]{\begin{tabular}{|c|c|c|c|c|}
	\hline
	Method & Training Data & Input & Output &  Error Analysis\\
	\hline 
	Generative Fitting Pipeline & grid of synthetic spectra &theoretical models & interpolated model spectra \& label predictions & Bayesian MCMC \\ 
	Parametric Random Forest & SDSS spectra with known labels & line profile parameters  & label predictions  & Bootstrap Resampling \\
	\hline 
	Traditional Approach & N/A & theoretical models & label predictions & $\chi^2$ statistics \\ 
	\hline 
	\end{tabular} }
	\centering
	\caption{Summary of methods presented in this work, along with the traditional approach for comparison.} 
	\label{table:methods}
\end{table*} 
 
 Selecting a region around each Balmer line from H$\alpha$ to H$8$ on the un-normalized spectrum, we fit a composite model of a Voigt profile added to a linear function of wavelength. The Voigt profile is the convolution of a Gaussian and Lorentzian distribution, and represents well the pressure-broadened wings of the Balmer lines \citep{Tremblay2009}. We divide the region around each Balmer line by the linear component of this composite fit to derive the continuum normalized spectrum for that line. 
 
 Our entire continuum-normalized spectrum is therefore discontinuous and made up of six distinct regions with some gaps in between. These regions are selected by cropping the spectrum at a fixed distance away from each Balmer line -- 300 \rm \AA\ for H$\alpha$, 200 \rm \AA\ for H$\beta$, 120 \rm \AA\ for H$\gamma$, 70 \rm \AA\ for H$\delta$, 50 \rm \AA\ for H$\epsilon$ and 25 \rm \AA\ for H$8$. Higher-order Balmer lines are too weak to be detected for large parts of our label space, and are hence omitted from our main fitting tool. The regions are selected to be large enough to include a sufficient portion of the surrounding continuum, whilst excluding nearby Balmer lines and strong telluric features. The resulting continuum-normalized spectrum is therefore unbiased relative to the depth and width of the absorption lines, and is also independent of the shape of the overall underlying black-body continuum \citep{Bergeron1995}. Normalizing each line independently also accommodates cases where clean spectral data is only available for some subset of the Balmer lines. 
 
 For helium-rich DB spectra, there are far more features on the spectrum and the above approach is not feasible. Instead we normalize the entire continuum at once. We pre-define a mask that excludes all known hydrogen and helium absorption lines on the optical spectrum, and fit a smoothed spline to the remaining pixels. We divide the observed spectrum by the fitted spline curve to obtain the continuum-normalized spectrum. This method is quite effective at normalizing the entire continuum to unity, albeit it suffers slightly on either end of the spectrum. If the spectrum is known to be DA, we prefer the above approach of normalizing each Balmer line individually. 
 
 All subsequent analysis in this work is performed on spectra continuum-normalized using the methods detailed above. It is also possible to independently constrain $T_{\text{eff}}$ from the black-body shape of the thermal continuum. Whilst directly deriving this constraint is outside the scope of this work due to numerous confounding factors like spectro-photometric calibration, the methods in Section \ref{synthetic} are designed to incorporate such prior knowledge of $T_{\text{eff}}$ if available.

\section{Methods}

In this section we summarize the analysis techniques implemented in {\sc wdtools} to measure the astrophysical properties of white dwarfs from their spectra. We also tabulate the inputs, outputs, and error analysis of each method in Table \ref{table:methods}.

\subsection{Generative Fitting Pipeline} \label{synthetic}
	
In our first method, we fit state-of-the-art atmospheric models and the resulting synthetic spectra to SDSS spectroscopic data. State of the art white dwarf synthetic spectra are based on theoretical models for pressure and thermal broadening to accurately reproduce the hydrogen Balmer lines as a function of the stellar labels. The codes to generate these spectra are not publicly available, but the methodology of one of the most popular codes is described in \cite{Koester2010}. These models incorporate the hydrogen Stark broadening with nonideal effects from \cite{Tremblay2009}, and assume local thermodynamic equilibrium. 

Around $1274$ DA (pure-hydrogen atmosphere) synthetic spectra are publicly available \footnote{\label{SVO} Courtesy the \href{http://svo2.cab.inta-csic.es/theory//newov2/index.php?models=koester2}{Spanish Virtual Observatory}.} that span $6,000 - 40,000$ K in $T_{\text{eff}}$ and 6.5 -- 9.5 dex in $\log{g}$, where $g$ is in CGS units of $\rm cm\ s^{-2}$. Additionally, 625 DB (pure-helium) synthetic spectra were provided to us (\citealt{KoesterPrivate}, private communication) spanning $6,000 - 40,000$ K in $T_{\text{eff}}$ and 7 -- 9 dex in $\log{g}$. In this work we test and discuss both the DA and DB models for completeness. However, the DB models are under restricted access, and we therefore do not include them in our accompanying software package. In future if these DB synthetic spectra become publicly available, we will add them to our software.

The usual fitting methodology is to fit the labeled model spectra to continuum-normalized absorption-line data and report the labels with the lowest $\chi^2$ error \citep{Kepler2019}. However, this approach has some drawbacks. The published model grid is sampled sparsely, requiring interpolation between label values and multiple function evaluations. As mentioned above, the codes to generate atmospheric models directly are not publicly available. Even if they were, modeling stellar atmospheres is quite computationally expensive and requires generating model grids in advance before the fitting procedure \citep{Koester2008}. \cite{Ting2018} propose neural networks as speedy, reliable interpolators that require few ab-initio models to generate high-quality synthetic spectra for main sequence stars. We present an adaptation of this approach using white dwarf theoretical models to speedily generate synthetic spectra in a way that can be incorporated into the fitting routine itself. 

We begin with the assumption that the stellar labels - $T_{\text{eff}}$ and $\log{g}$ - uniquely determine the pixels on a continuum-normalized spectrum \citep{Ness2015}. We could include radial velocity as a label here, but instead choose to leave it as a free nuisance parameter during the final fitting procedure. We pre-process the models by using linear interpolation to bring all model spectra onto an identical wavelength range in 0.5 \AA\ intervals, and then continuum-normalize them with the procedures described in Sec. \ref{contcorrection}. The model spectra are later brought down to the resolution of the observed spectrum using Gaussian convolution during the fitting step, and we therefore sample the ab-initio models finely for use with higher-resolution spectroscopy. We are happy to provide pre-trained neural networks that interpolate an even finer grid of wavelengths upon request.

The generative neural network takes in stellar labels and creates continuum-normalized synthetic spectra from them. The network's input layer consists of 2 neurons, corresponding to the 2 labels: $T_{\text{eff}}$ and $\log{g}$. The output layer has thousands of neurons, each corresponding to one pixel on the continuum-normalized spectrum. The aim of training the network is to predict an accurate normalized flux at every pixel given only the 2 labels at the start. The bulk of the network's thinking occurs in the so-called `hidden' layers between the input and the output. The goal of the training step is to find the optimal combination of weights and biases $w,b$ in the hidden layers to minimize the mean squared error between the flux predicted by the neural network and the true flux from the synthetic models. Through experimentation, we find that using 256 neurons in each of 3 hidden layers is optimal, allowing the neural network to fully learn the non-linear relationships between the labels and spectrum whilst minimizing interpolation artefacts.

\begin{figure}
 	\includegraphics[width = \columnwidth]{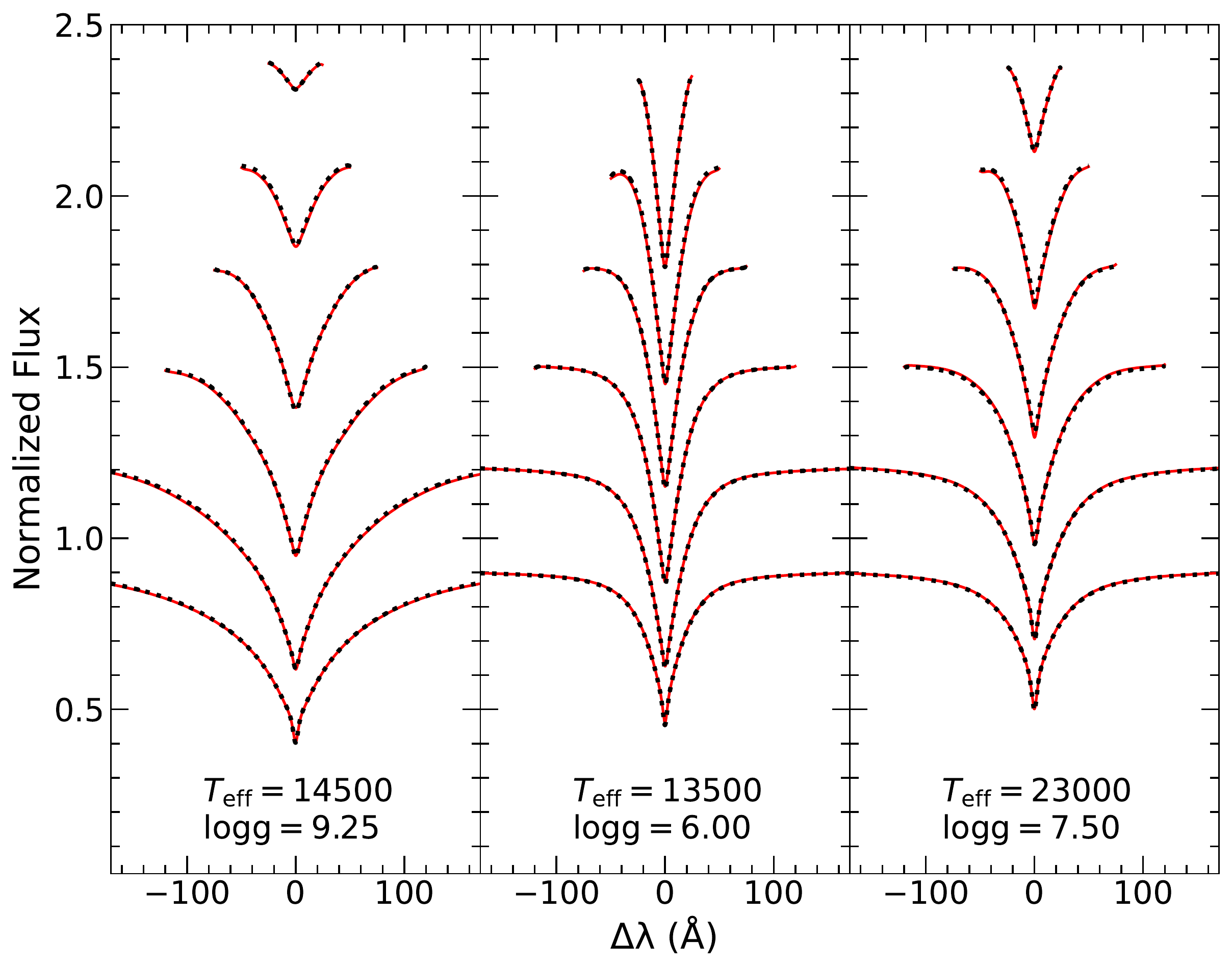}
 	\includegraphics[width=\columnwidth]{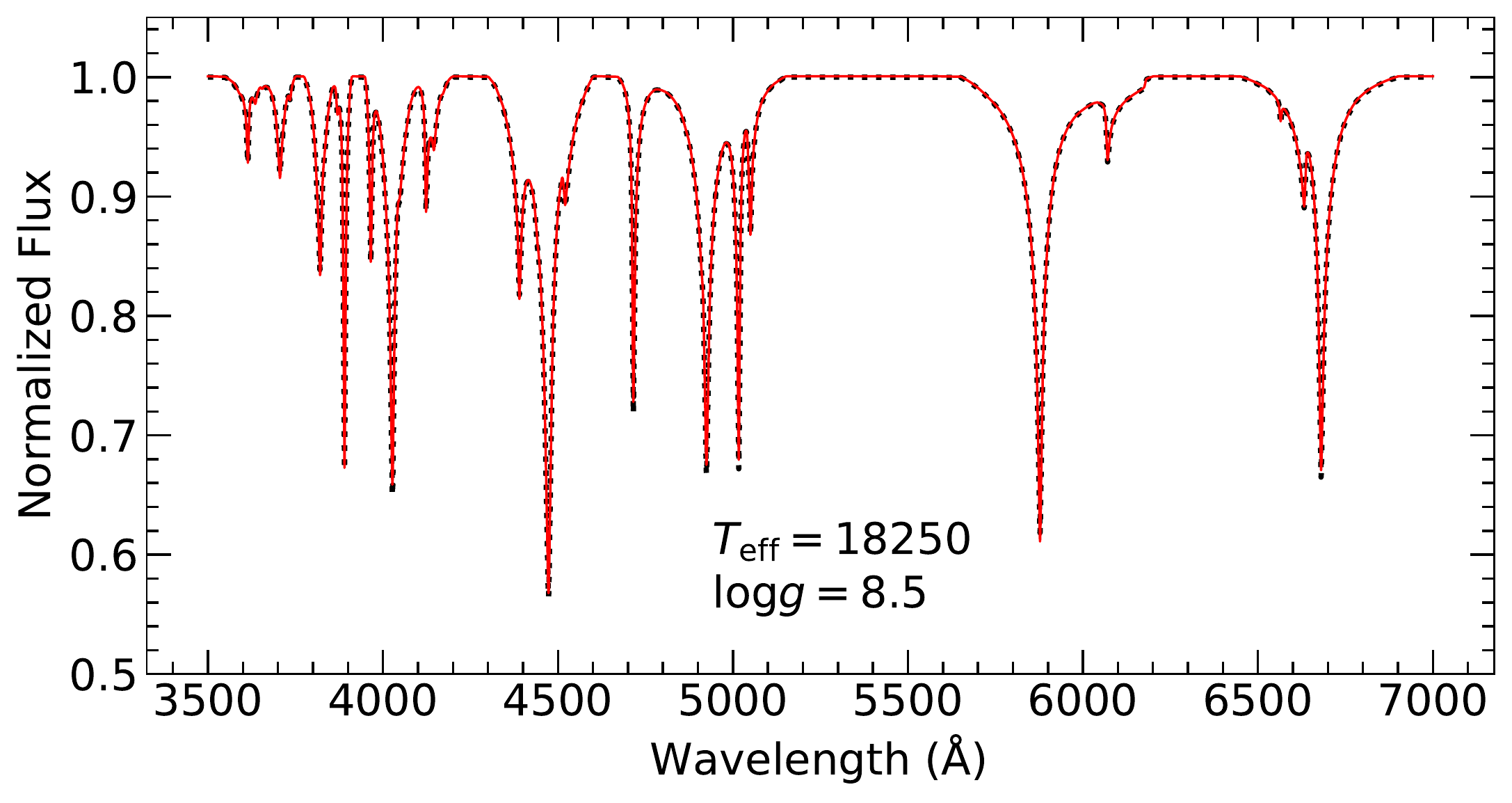}
	\caption{ The neural networks accurately reproduce synthetic DA (top) and DB (bottom) spectra from the validation set of labels unseen during training. In red is the spectrum generated by the neural network, and in black is the true synthetic spectrum for those labels. For the DA stars the continuum-normalized Balmer lines are vertically displaced for clarity and are arranged from H$\alpha$ at the bottom to H$8$ at the top. For DB stars the entire continuum-normalized optical spectrum is interpolated.}
	\label{fig:genspec} \label{fig:dbpred} 
\end{figure}

The neural network is relatively inexpensive to train, taking a few hours of CPU time on a regular laptop to train itself on the continuum-normalized model spectra. Once trained, the network generates synthetic spectra from labels extremely quickly on the order of 500 spectra per second. With any machine learning technique that trains itself on data, it is good practice to test the model on data outside the original training set. This ensures that the model has correctly learned the underlying physics rather than `over-fit' the training dataset. We therefore randomly reserve 1 percent of the model spectra for validation, and keep them out of the training procedure. Testing the interpolation error on these validation labels (unseen during training) we obtain a relative error of under 0.01 in normalized flux units across all pixels (Figure \ref{fig:genspec}).

\begin{figure}
    \centering
    \includegraphics[width=\columnwidth]{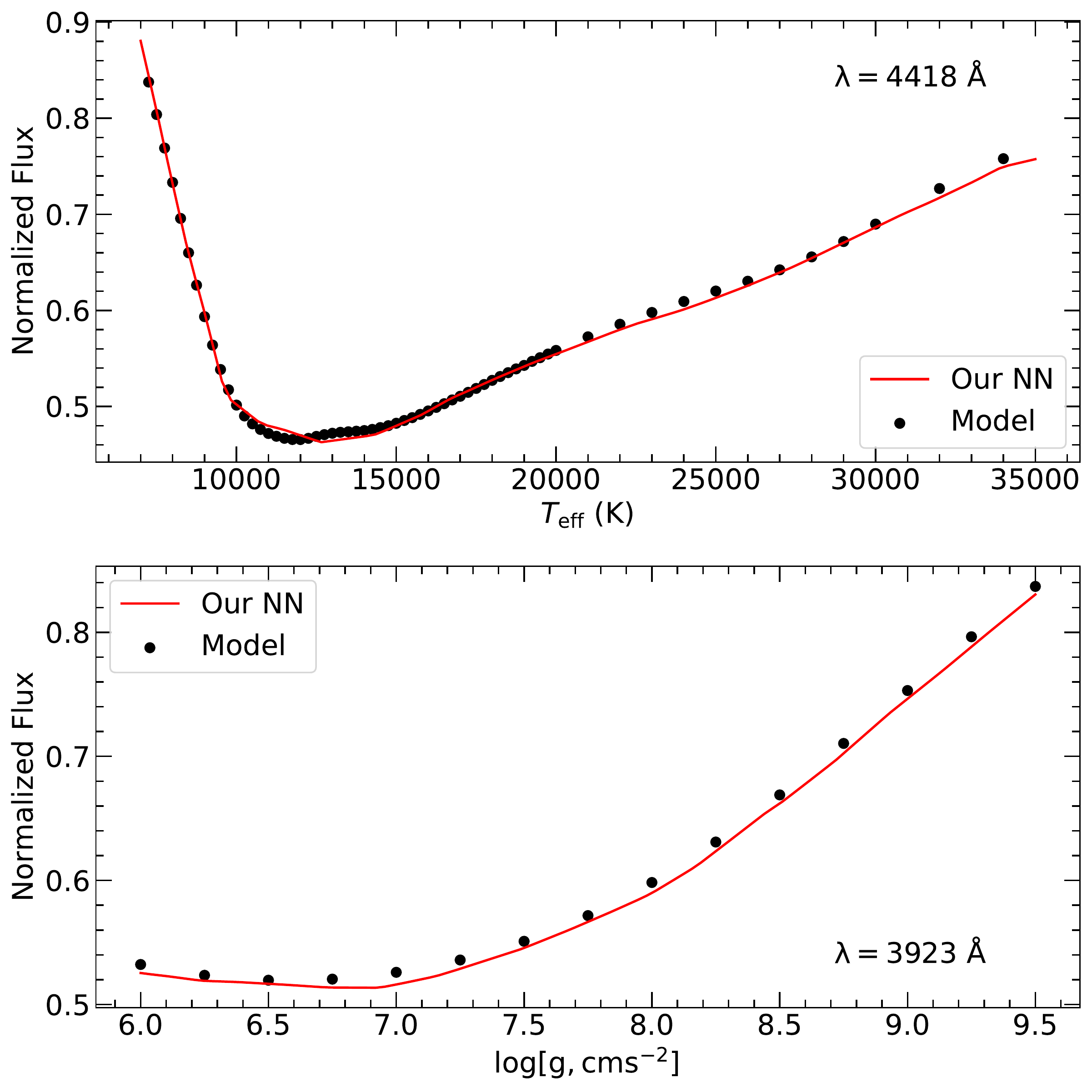}
    \caption{The interpolation accuracy of our DA generative neural network in label space, for two random pixels on the continuum-normalized spectrum. The neural network reproduces the expected non-linear behaviour without any signs of interpolation artefacts or over-fitting. The interpolation quality is similarly good across all pixels.}
    \label{fig:interp_labels}
\end{figure}

We differ from the architecture of the {\sc payne} described in \cite{Ting2018} by having a broader neural network with 2 hidden layers of 256 neurons each, rather than 10 neurons each. Additionally, \cite{Ting2018} trained a different neural network for each spectral pixel, but we train a single network for the entire spectrum. This incorporates the information and mapping between adjacent pixels. Whilst our neural network has more free parameters, it is capable of learning more non-linear relationships, resulting in an reduction in percent flux error of $\sim 10$ percent compared to if we used \cite{Ting2018}'s architecture instead. Additionally, since we use a single neural network, our training time is also much shorter \citep{Leung2018}. We find that using a broader neural network architecture does not produce overfitting artefacts - the interpolation is smooth across both stellar labels (Figure \ref{fig:interp_labels}). Newer versions of the {\sc payne} developed by \cite{Ting2018} also incorporate similar improvements in their architecture.

Armed with the ability to generate our own high-fidelity synthetic spectra very quickly, we are not limited to the original grid of model spectra. We can now create synthetic spectra `on the fly' during the fitting process. With unlimited synthetic spectra in hand, we proceed to construct a fitting pipeline that utilizes the neural network. Our goal is to fit for $T_{\text{eff}}$ and $\log{g}$ (hereafter $T$ and $l$ for brevity) on the continuum-normalized Balmer absorption lines. In the Bayesian framework, the posterior distribution of the stellar labels $f(T,l |\ X)$ of a single star is given by 
\begin{equation}
    f(T, l | X) = \frac{f_{p}(T,l) \cdot f(X | T, l)}{f(X)}
\end{equation} where $X$ is the data (continuum-normalized spectrum) and we define $f_{p}(T,l)$ is our prior knowledge of the stellar labels. By default, we use a uniform prior
\begin{equation}
    f_{p}(T,l) = {\mathbbm{1}}_{\left[6000,80000\right]}(T) \cdot  {\mathbbm{1}}_{\left[ 6.5,9.5\right]}(l)
\end{equation}
where $\mathbbm{1}$ denotes the indicator function on the stated range. This range is chosen to match the label range of the model spectra used to train the neural network, to prevent extrapolation errors. We include the ability to replace the uniform prior on $T_{\text{eff}}$ with a Gaussian prior if another constraint on $T_{\text{eff}}$ from photometry or black-body continuum is available. 

The purpose of the main fitting step is to determine the empirical distribution $\frac{f(X | T, l)}{f(X)}$. Since our fitting approach relies on the discrete sampling (in label space) of synthetic spectra, it is especially suited to fitting via Markov Chain Monte Carlo (MCMC) algorithms. We utilize the {\sc emcee} sampler for our MCMC procedure \citep{Foreman-Mackey2013}. {\sc emcee} samples the posterior distribution of the stellar labels using an affine-invariant version of the Metropolis-Hastings algorithm \citep{Goodman2010} on a defined likelihood. We assume a chi-square empirical likelihood \begin{equation}\label{chilik}
    \frac{f(X | T, l)}{f(X)} \sim\ \chi^2_{n}
\end{equation}
which is standard for curve fitting in astrophysics \citep{Bevington1969}. The number of degrees of freedom $n$ of $\chi^2_{n}$ is nominally defined as the number of pixels used minus the number of parameters in the fit. 

We include the radial velocity of the star as a `nuisance parameter' in the MCMC fit, marginalizing over it. This is preferred over correcting the spectrum to some fixed rest-frame with a single radial velocity measurement, as these measurements are susceptible to significant biases due to temperature and surface gravity \citep{Halenka2015}. Additionally for DB stars it is non-trivial to measure an accurate radial velocity due to the lack of unblended symmetric absorption lines, so leaving it as a free parameter is preferred. However, if the sole goal is to derive high-precision radial velocity measurements from spectra, a simpler and model-independent independent approach like fitting Gaussian profiles to a small region around the H$\alpha$ absorption line may be preferred \citep{Joyce2018}. This prevents any systematic shortcomings in the atmospheric models from biasing the radial velocity measurement.

We begin our fitting procedure for each star with two initial fitting steps to evaluate the `hot' and `cold' solutions for the star. For the cold solution, we restrict $6000 < T_{\rm eff} \leq 15000$ and for the hot solution we restrict $15000 < T_{\rm eff} < 40000$, allowing all other parameters to vary freely. We use the differential evolution optimization algorithm \citep{Storn1997} in {\sc lmfit} to minimize Equation \ref{chilik} and select the solution with the highest likelihood (lowest $\chi^2$). For the main MCMC fitting procedure, we initialize independent `walkers' in a tight Gaussian ball around the maximum likelihood region of parameter space derived by the initial fit. The walkers explore the parameter space for a burn-in period, during which they spread out and settle into uncorrelated samples from the posterior distribution. Finally we reset the sampler and sample the posterior label distribution $ f(T, l\ | X)$ for a number of steps, from which we compute the maximum likelihood (minumum $\chi^2$) stellar labels as well as label uncertainties using the standard deviation of each posterior distribution. This entire procedure entails upwards of 100,000 likelihood calls depending on the number of posterior samples required, yet takes less than a minute of CPU time per star on a regular laptop due to the speed of the generative neural network. 

In this section, we have demonstrated how even in the absence of access to a finely spaced grid of models, we can use machine learning techniques to interpolate high quality models for both DA and DB white dwarfs. We discuss the quality and limitations of this technique further in Section \ref{results}.  

\subsection{Parametric Random Forest} \label{rf}

In our second technique, we seek to improve traditional methods of parametrically fitting absorption line profiles to derive atmospheric parameters. DA white dwarfs are characterized by strong hydrogen absorption lines. A standard approach to analyze spectral absorption lines is the use of line profiles to quantify line parameters like centroid, width, and scale. White dwarf absorption lines are formed over a range of depths and pressures, and can be modelled by the convolution between a symmetric Voigt profile and an asymmetric Stark profile \citep{Tremblay2009,Tremblay2019}. We find that including asymmetric information adds complexity without signfificantly improving the accuracy of our model, and we therefore approximate the white dwarf line profiles using the symmetric Voigt distribution. 

We use Python's {\sc LMFIT} \citep{Newville2014} to fit Voigt profiles to the H$\alpha$, H$\beta$, H$\gamma$, and H$\delta$ absorption lines, performing a $\chi^2$ minimization each time. We leave out higher-order lines since they aren't well resolved for a lot of stars in our sample, and we want to ensure accurate profile fits. Our fitting procedure yields 2 parameters for each line: full-width at half-maximum (FWHM) and amplitude. The line centroid is allowed to vary freely during the fit to account for radial velocity and is discarded as a nuisance parameter. These line profiles do not perfectly fit the entire absorption line -- they are designed to fit the wings well, at the cost of the absorption line core. This is because most of the physical information about the stellar labels is encoded in the wings, not the core \citep{Halenka2015}. Additionally, the line core is usually resolution-limited, having a width of $\sigma_v \simeq 30\ \rm km\ s^{-1}$ compared to the SDSS resolution of $\sigma_v \simeq 60\ \rm km\ s^{-1}$ at H$\rm \alpha$. 

\begin{figure}
	\includegraphics[width = \columnwidth]{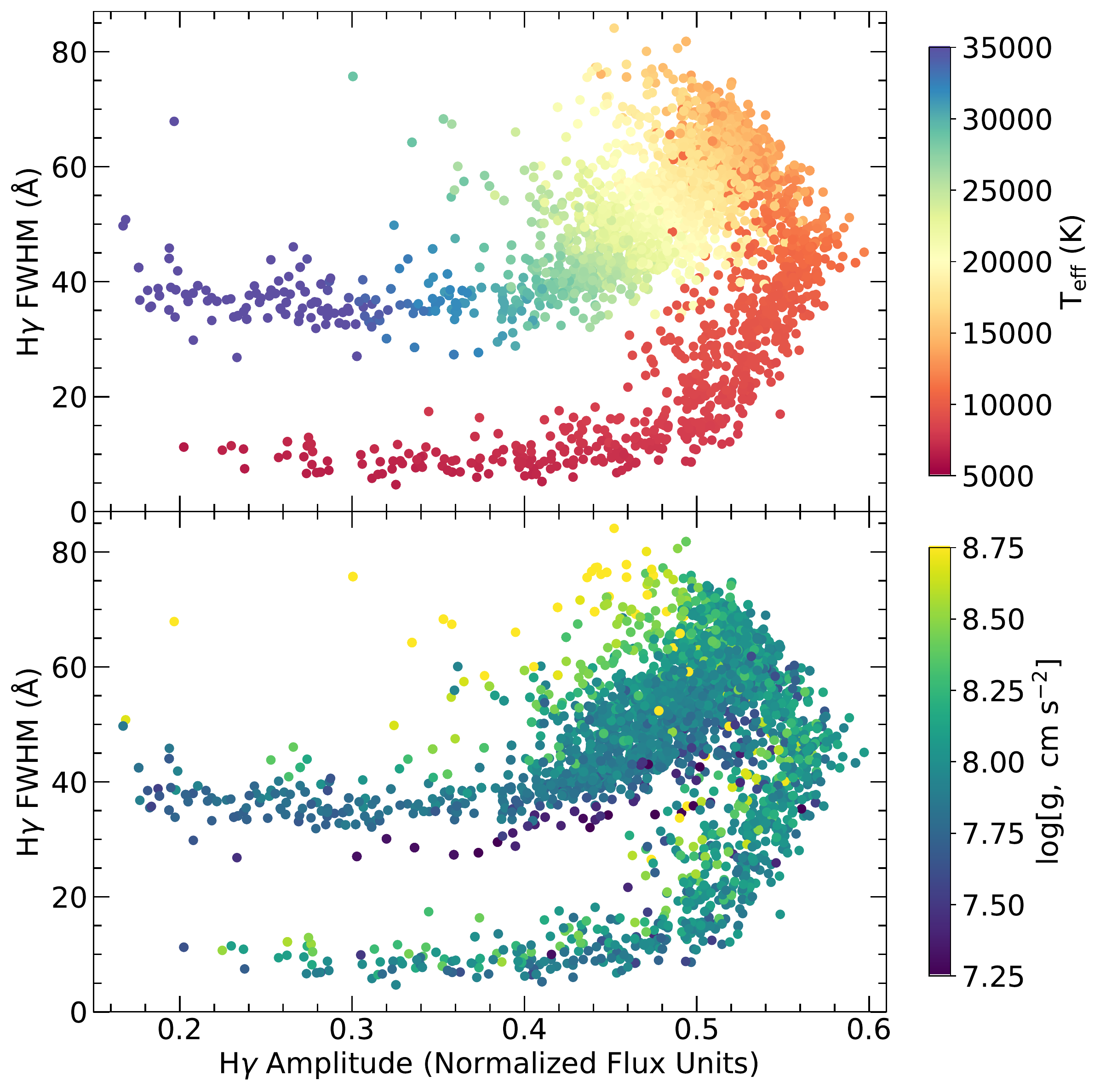}
	\caption{Visualizing the `phase space' of the H$\gamma$ absorption line using two parameters from our fitted profiles, the full-width at half-maximum and the line amplitude. There are intricate two-dimensional structures and trends in this phase space as a function of the stellar labels $T_{\mathrm{eff}}$ and $\log{g}$.}
	\label{fig:phasespace} 
\end{figure}

\begin{figure}
	\includegraphics[width = \columnwidth]{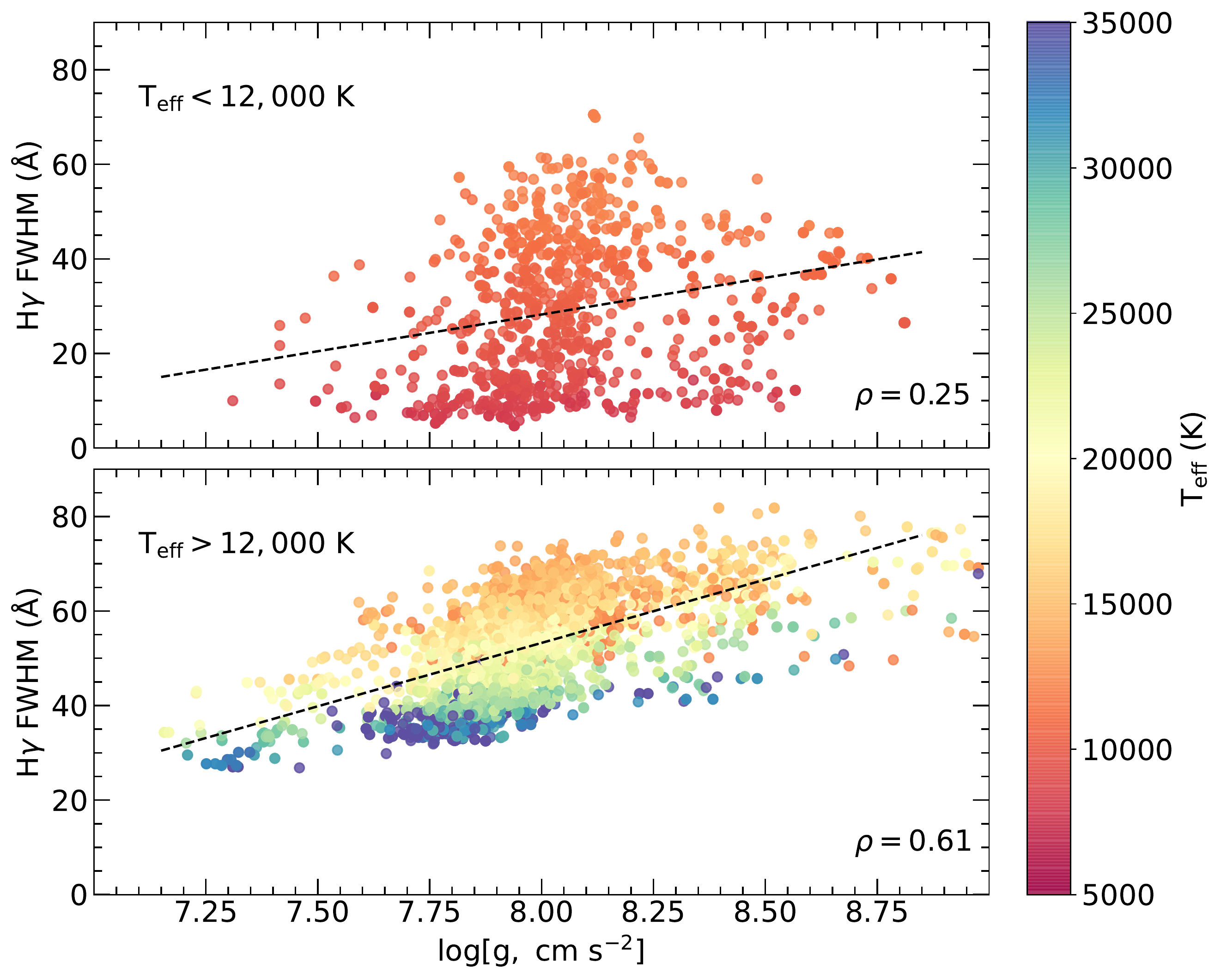}
	\caption{The width of H$\gamma$ is more positively correlated with $\log{g}$ for stars with ${T_{\text{eff}}} > 12,000$ K and therefore carries more information about $\log{g}$ at high temperatures. Overlaid in dashed black is a linear regression fit, and $\rho$ denotes the Pearson correlation coefficient between $\log{g}$ and the H$\gamma$ FWHM.}
	\label{fig:logg_trend} 
\end{figure}

Absorption lines are known to predictably vary with changes in $T_{\text{eff}}$ and $\log{g}$, due to effects like pressure broadening. Figure \ref{fig:phasespace} demonstrates the relation between our derived width and depth for the H$\gamma$ line colored by known spectroscopic $T_{\text{eff}}$ and $\log{g}$ labels from \cite{Tremblay2019}. We select H$\gamma$ for this figure since it is the most sensitive to these labels -- H$\alpha$, H$\beta$ and H$\delta$ exhibit similar relations. It is clear from Figure \ref{fig:phasespace} that it is intractable to distinguish $T_{\text{eff}}$ and $\log{g}$ from any single line parameter; a particular value of FWHM, for example, would correspond to multiple sets of stellar labels. Additionally, the absorption line width is more sensitive to $\log{g}$ at higher temperatures than at temperatures below $\sim 12000$ K (Figure \ref{fig:logg_trend}). The goal of our multi-parameter approach is to combine the information of the parameters across all lines to distinguish and fit for $T_{\text{eff}}$ and $\log{g}$. 

We use a multivariate random forest regression model to map line parameters to stellar labels. Random forests \citep{Breiman2001} have become a popular tool in astronomy to learn non-linear relations between features, in both supervised \citep{Richards2011} and unsupervised \citep{Reis2019} learning environments. A random forest is composed of logical structures called decision trees. Each decision tree maps the input feature space (in our case the 8 absorption line summary statistics) to continuous values in the target variable space (in our case the 2 stellar labels). A tree begins with a root node comprising of the entire input feature space - in our case the feature space is composed of 2 parameters from each of four Balmer lines, for a total of 8 features. It then splits into several nodes, each of which represents a different subset of the feature space. It continues this recursive partitioning until all leaves originating from a particular node share (within a tolerance) the same value of the target variable. 

Whilst decision trees on their own are powerful tools for regression, their predictive power is hampered by their tendency to over-fit training data. This results in a poorly generalized model which fails to make meaningful predictions outside the dataset it was trained on. For some applications, it is possible to fix this problem by combining an ensemble of decision trees through bootstrap aggregation (`bagging'; \citealt{Breiman2001}). Each decision tree is trained on a bootstrap sample of the training data rather than the entire dataset, and the results of all trees in the ensemble are averaged to produce the final regression output. Bagging limits over-fitting and reduces the variance of the regression. 

One challenge with random forests is that they do not natively quantify uncertainties or errors. We implement our own method to produce prediction uncertainties based on \cite{Coulston2016}. We construct 100 bootstrap resamples of the stars in our training sample, sampling with replacement. We train an ensemble of 100 random forest models, one on each of the above bootstrapped samples. We make predictions using this entire ensemble, and use the variance of predictions across all component decision trees to define a test statistic as defined in \cite{Coulston2016}, and subsequently obtain 1-$\sigma$ uncertainties for our fitted labels.  Since each random forest in the ensemble is trained on an overlapping but different set of stars, the aggregate predictions of the ensemble capture the predictive uncertainty of the model \citep{Coulston2016}. Additionally, the variance across all the individual decision trees can capture the epistemic and noise uncertainty in the data, since each tree can be thought of as a `weak learner'. When testing our predictions on a validation set of stars, we find that $\sim 60$ percent of our predicted labels lie within 1-$\sigma$ of the ground truth values obtained by prior studies, and $\sim 85$ percent lie within 2-$\sigma$. This indicates that our uncertainty estimation procedure slightly underestimates the errors, but is otherwise in good agreement with Gaussian statistics. 

Another possible challenge is the non-uniform distribution of our training sample in label space. In particular, the $\log{g}$ distribution is sharply peaked around $\log{g} \sim 8$ dex. Since most training stars have labels around this value, the random forest algorithm will tend towards predicting values closer to the mean - this worsens the prediction accuracy near the tails of the distribution. We combat this problem during the ensemble resampling step. We fit a two-dimensional Gaussian to the density distribution of the objects in the training label space (over $T_{\text{eff}}$ and $\log{g}$) to obtain the kernel density estimate (KDE) for the sample. We then assign each star a weight equivalent to the reciprocal of the KDE's probability density value for that star's labels. We sample our stars with replacement using these weights during the ensemble resampling. Therefore, stars near the tails of the label distributions are favored, and the distribution of labels used to train each random forest is more uniform. 

We use the {\sc SciKit-Learn} library in Python to construct our random forest models with 25 decision trees each \citep{Pedregosa2011}. There are a total of 8 features derived from each stellar spectrum - amplitude and FWHM for H$\alpha$, H$\beta$, H$\gamma$, and H$\delta$ respectively. These features are used to predict the two stellar labels ($T_{\rm eff}$ and $\log{g}$) simultaneously. Our ensemble model is fast to train, on the order of a few seconds for 5000 stars; predicting labels for thousands of stars at once likewise takes a couple of seconds on a regular laptop computer. 

In this section we have described a way to combine parametric information from all the absorption lines on a white dwarf spectrum to make inferences about stellar labels. This should result in a less noisy model compared to direct pixel methods, since the parametric line fit incorporates much of the noise in the spectrum flux itself. Given a large sample of spectra of which some subset has accurate labels, this method is powerful for inferring labels for the entire sample. This is a form of `transfer learning', and we have previously used it to derive spectroscopic $\log{g}$ for a large sample of SDSS white dwarfs \citep{Chandra2020}. We evaluate this method on SDSS spectra in Section \ref{results}.

\subsection{Convolutional Neural Networks} \label{baynn}

In recent years interest in convolutional neural networks (CNNs) has grown, primarily in the fields of image recognition and visual artificial intelligence. \cite{Leung2018} describe convolutional neural networks with intrinsic uncertainty estimation to infer abundances from APOGEE spectra. We adapt their architecture to develop a Bayesian convolutional neural network that is trained to predict $\rm T_{\rm eff}$ and $\log{g}$ from SDSS white dwarf spectra directly.  The `true' labels used to train this model are taken from the spectroscopic fits of \cite{Tremblay2019}. We use our CNN to simultaneously incorporate all continuum-normalized spectral pixels into a regression model to predict stellar labels. 

In practice, the convolutional neural network achieves label prediction accuracy similar to the random forest method described in Section \ref{rf}. However, this kind of `black-box' prediction algorithm suffers badly from a lack of interpretability. For example, since the training set of stars are all pure-hydrogen DA stars, the neural network makes unrealistic predictions when faced with a non-DA star. We find that whilst the data-driven neural network prediction models work well for applications like \cite{Leung2018} where several element abundances must be constrained from a complex and blended spectrum, the simpler and more interpretable methods of Section \ref{synthetic} and \ref{rf} work better for our data. We therefore include this discussion of CNNs here for completeness, but do not consider them further in our results or software package. 
 
\section{Results}\label{results}

\subsection{DA White Dwarfs from the Sloan Digital Sky Survey}\label{sdssresults}
	
\cite{Tremblay2019} derived spectroscopic $T_{\text{eff}}$ and $\log{g}$ for a sample of 5327 DA white dwarfs using their latest atmospheric model codes. These models have been shown to be consistent with results obtained by other groups and photometry, with typical root mean square (RMS) discrepancy between stellar labels derived by different methods being on the order of $500$ K in $T_{\rm eff}$ and 0.1 dex in $\log{g}$. We use the stellar labels from \cite{Tremblay2019} as a `ground truth' sample to test the methods described in this work. For consistent model comparison with the random forest regression, we divide the sample into a training set of 4011 stars (75 percent) and a testing set of 1316 stars (25 percent). 
	
\begin{figure*}
	\includegraphics[width=\textwidth]{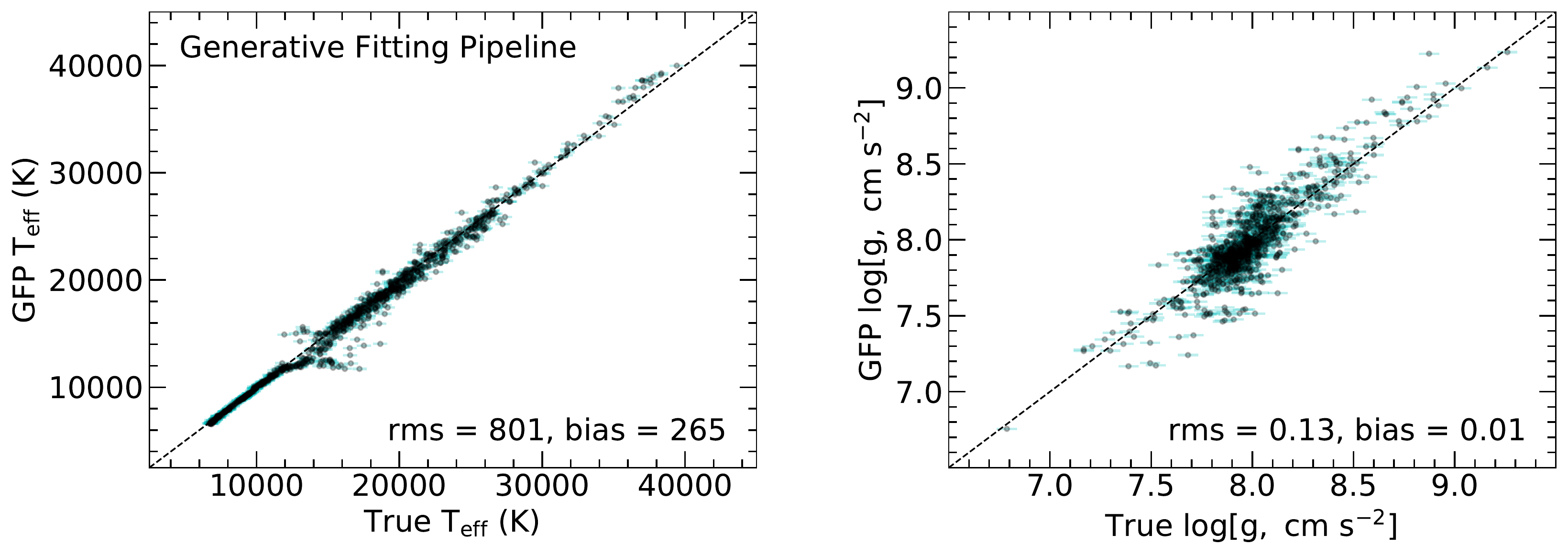}
	\includegraphics[width=\textwidth]{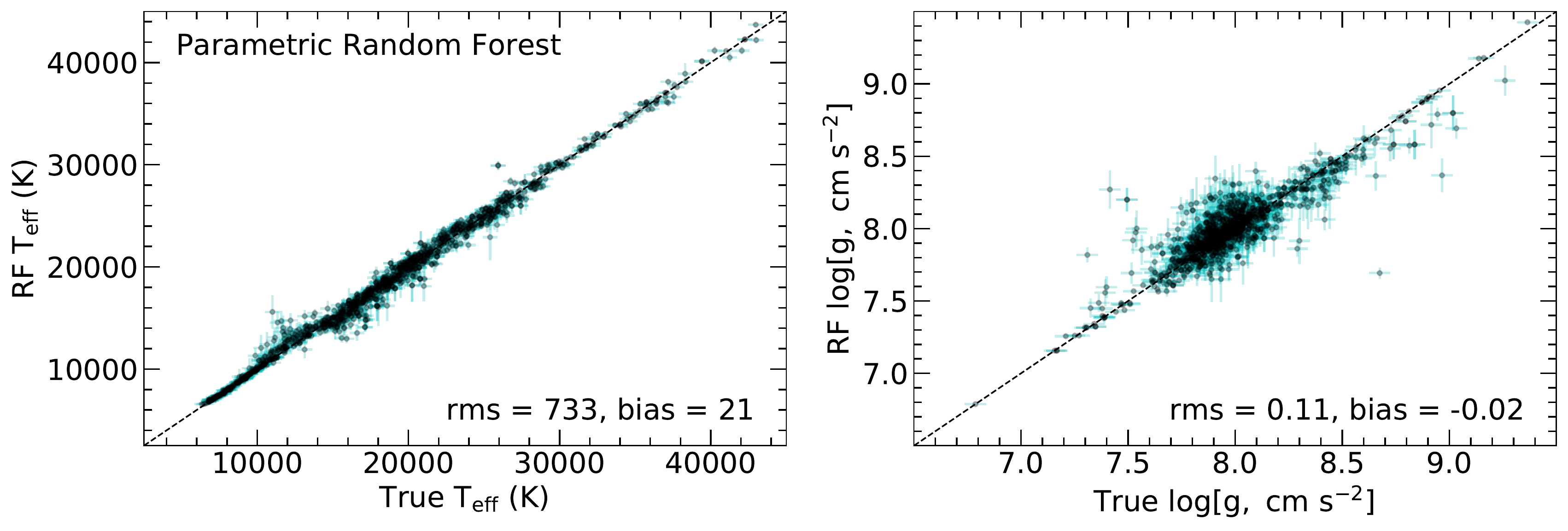}
	\caption{Summary of the prediction accuracy of the generative fitting pipeline (top) and random forest model (bottom) for both $T_{\text{eff}}$ (left) and $\log{g}$ (right), comparing our labels for a validation set of 1316 DA SDSS spectra to `true' labels derived from an earlier spectroscopic study \citep{Tremblay2019}. 1-$\sigma$ error bars as reported by the respective methods are overlaid in cyan.}
	\label{fig:nnpred} 
\end{figure*}

We run our generative fitting pipeline on all 1316 spectra in the testing set, obtaining posterior label distributions in a minute per star on average. These fits only consider the information on the spectrum. In practice, it may be possible and desirable to combine spectroscopic data with photometry and obtain a prior on the temperature, which may improve the spectroscopic fit. Our MCMC fitting approach almost always finds the global $\chi^2$ minimum, since a choice between hot and cold solutions on the basis of $\chi^2$ is made during our pre burn-in phase where we select the highest probability region of the parameter space. Regardless we recommend always using photometry, if available, as an additional constraint on $T_{\text{eff}}$. We present our fitted labels compared to \cite{Tremblay2019} in Figure \ref{fig:nnpred} (top).

We train the bootstrap random forest ensemble on the training set of 4011 stars and make predictions with uncertainties on the test set of 1316 spectra. Our set of training features consists of line summaries from H$\alpha$-H$\delta$, since these lines are very well resolved on all stars in our training set of stars. We compute uncertainties as described in Section \ref{rf} and present our predictions compared to \cite{Tremblay2019} in Figure \ref{fig:nnpred} (bottom). 

We can use the uncertainties reported by our methods to remove grossly under-determined labels -- we reject labels with uncertainty in $T_{\text{eff}} \geq 2500$ K and uncertainty in $\log{g} \geq 0.5$ dex. These stars also have systematically higher $\chi_r^2$, although we find that $\chi^2_r$ alone is usually not sufficient to reject spurious fits (see Section \ref{outliers}). Upon inspecting these stars further, we find that they usually have some deviation from the model assumptions, like weak magnetism or a very low signal-to-noise ratio. We compare our derived labels to those from \cite{Tremblay2019} in Figure \ref{fig:nnpred} with the minimal selection cuts on reported error described above. Both techniques recover $T_{\text{eff}}$ within $\sim$ 1000 K for all stars, and $\log{g}$ within $\sim$ 0.12 dex. 

\subsection{DB White Dwarfs from the Sloan Digital Sky Survey}

\cite{Koester2015} identified 1000 SDSS spectra with a DB classification and fit them with their atmospheric models. We fit these spectra with our generative fitting pipeline. The $\log{g}$ constraint is not particularly specific since the atmospheric models are not very sensitive to variations in $\log{g}$ across all temperatures. Furthermore, \cite{Koester2015} noted that $\log{g}$ is over-estimated for stars with $T_{\text{eff}} \lesssim 15000$ K, which we find with our labels as well. This effect represents a well-known systematic shortcoming in our current understanding of white dwarf atmospheres. We therefore ignore $\log{g}$ labels for stars with $T_{\text{eff}} <$ 15000 K. 

We recover $\rm T_{\rm eff}$ within $\sim 1200$ K across the entire range of temperatures (6000 K - 40000 K) and $\log{g}$ within 0.25 dex for stars with $T_{\text{eff}} > 15000$ K. The root mean square discrepancy between our temperatures and the prior work is $632$ K if we only consider stars with $T_{\text{eff}} < 22000$ K (90 percent of the sample). As with \cite{Koester2015}, we find a gap in the fitted temperatures between 22000-24000 K and a subsequent degradation of the quality of fits for higher temperatures. Overall, our label predictions are consistent with the fundamental parameter accuracy from \cite{Tremblay2019}, who directly compared the \cite{Koester2015} spectroscopic labels to photometric labels from Gaia. 

As mentioned above, the DB models provided to us are under restricted access and hence omitted from our accompanying package until they become publicly available. However, we have demonstrated that given a grid of DB models, our methods produce results consistent with a sample of SDSS DB stars from \cite{Koester2015}.

\subsection{DA White Dwarfs from VLT--UVES}

As a part of the ESO Supernova Ia Progenitor Survey (SPY), \cite{Koester2009} obtained high-resolution spectra of 615 DA white dwarfs with the UV-Visual Echelle Spectrograph (UVES; \citealt{Dekker2000}) of the Very Large Telescope (VLT). The spectral resolution at H$\alpha$ is around $R = 18 500$ with per-pixel signal-to-noise $\geq 15$ for most spectra. \cite{Koester2009} derived atmospheric parameters for these white dwarfs using their models and $\chi^2$ minimization. Since then their models have changed considerably, especially with the inclusion of effects from \cite{Tremblay2009}. Our generative neural network is trained on synthetic spectra from the latest models that include these additional effects. We therefore revisit the spectroscopic catalog of \cite{Koester2009} using our generative fitting pipeline. 

We continuum-normalize the H$\gamma$, H$\delta$, H$\epsilon$, and H8 spectra from cross-disperser \#2 of UVES. We do not include the H$\alpha$ and H$\beta$ lines from cross-disperser \#3 due to their lower signal-to-noise and the presence of noisy artefacts. We use the generative fitting pipeline described in Section \ref{synthetic} to infer $T_{\text{eff}}$ and $\log{g}$ from the continuum-normalized spectra. 

Comparing fitted labels from our generative fitting pipeline to prior labels from \cite{Koester2009}, we find a root-mean-square discrepancy of $1500$ K in $T_{\text{eff}}$ and 0.2 dex in $\log{g}$. For stars above $20,000$ K, there is an interesting systematic bias of $\sim 1000 K$ between our labels and those from \cite{Koester2009}, albeit with low variance. Overall we predict systematically lower $T_{\rm eff}$ by $1000$ K and systematically higher $\log{g}$ by 0.2 dex. We attribute this bias to differences in the models used by our software and those used by \cite{Koester2009} over a decade ago, since the newer models include significant new physics from \cite{Tremblay2009} and other theoretical studies. 

We also use the random forest method of Section \ref{rf}, pre-trained on line summaries from the H$\delta$ and H$\gamma$ lines of 5327 DA white dwarfs from the SDSS as described in Section \ref{sdssresults}. Our software tool includes the capability to easily switch the model to any subset of the first 4 Balmer lines, and we test its applications beyond SDSS spectra here. Fitting the SPY sample with our random forest model, we recover the labels from \cite{Koester2009} to within 1000 K in $T_{\rm eff}$ and 0.25 dex in $\log{g}$ (RMSE). This method systematically overestimates $T_{\rm eff}$ by 200 K and $\log{g}$ by 0.1 dex, an improvement over the generative fitting pipeline. Our interpretation of this improvement is that since the random forest relies on gross line summaries (FWHM and amplitude) of the absorption lines, it is less sensitive to the intricate differences in physics between the older atmospheric models and the newer ones. 

\subsection{Finding Outlier Spectra}\label{outliers}

Apart from inferring stellar labels, the generative fitting pipeline (Sec. \ref{synthetic}) can be used to find interesting spectra that violate the assumptions of the atmospheric model. For DA white dwarfs our assumption is a pure hydrogen atmosphere and no magnetic field, with no metal pollution. \cite{Kepler2019} provide a catalog of over 35,000 SDSS white dwarfs with visually confirmed spectroscopic classifications. We fit pure-hydrogen models to a random subset of the spectra in this catalogue and compute the reduced $\chi^2_r$ goodness-of-fit statistic ($\chi^2$ per degree of freedom; \citealt{Bevington1969}). We find that for all spectra classified as pure-hydrogen (DA) by \cite{Kepler2019}, the $\chi^2_r \sim 1$ as expected, usually slightly lower. However, for all other spectral types, the $\chi^2_r$ values are systematically higher, including for DAH (magnetic) white dwarfs. 

However, the $\chi^2_r$ statistic is not sensitive enough for complete outlier removal by itself. That is, several spectra that are not pure-hydrogen also have $\chi^2_r \sim 1$ due to overestimated noise or other effects. The 99th percentile $\chi^2_r$ among the pure-DA stars is 1.25. By applying a selection cut of $\chi^2_r < 1.25$ to the fitted spectra, we retain completeness of 99 percent the true-positive DA stars and get rid of $\sim 50$ percent of the contaminant spectra. The remaining contaminant spectra cannot be differentiated on the basis of $\chi^2_r$ alone. 

\begin{figure}
    \centering
    \includegraphics[width=\columnwidth]{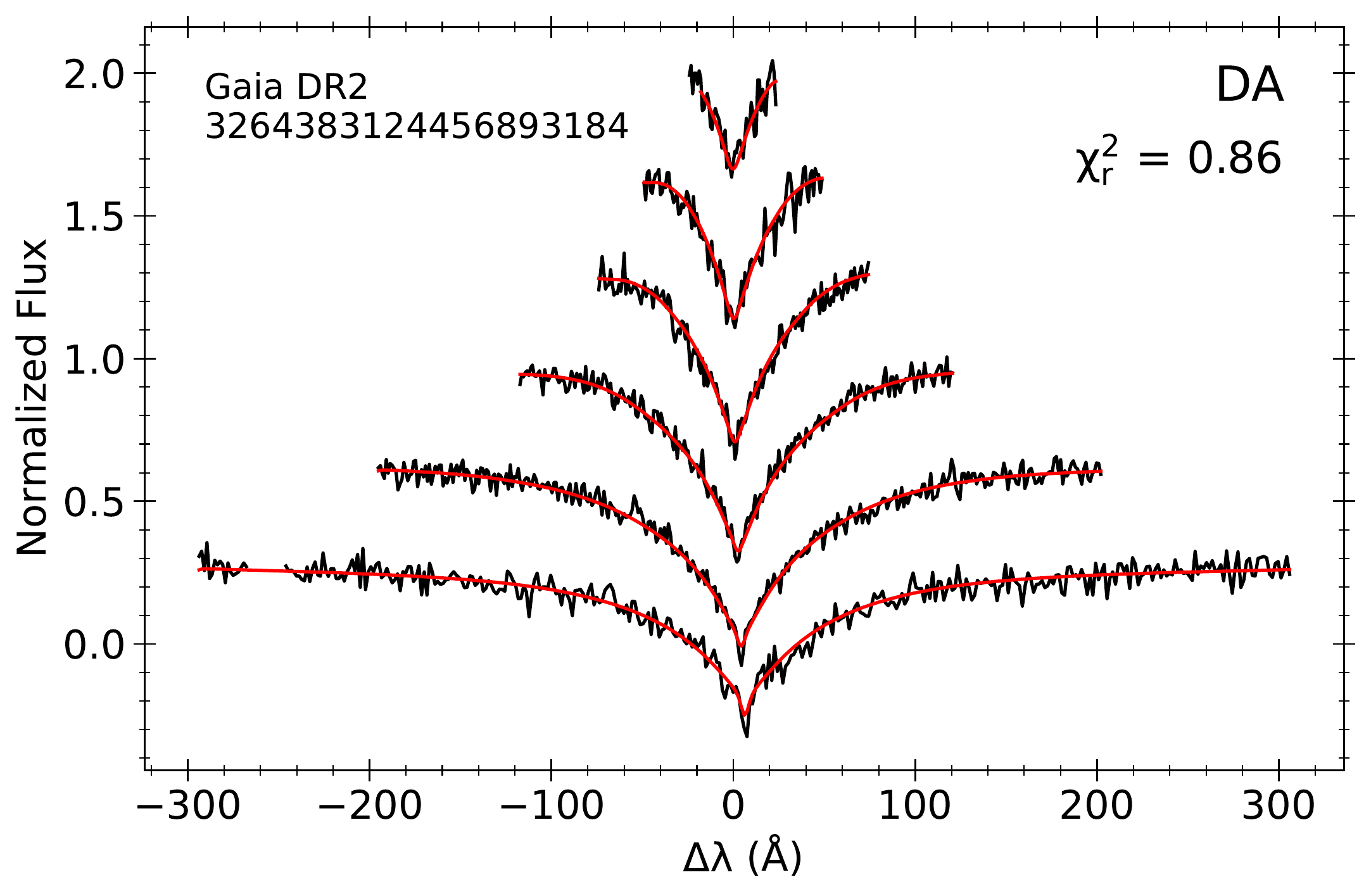}
    \includegraphics[width=\columnwidth]{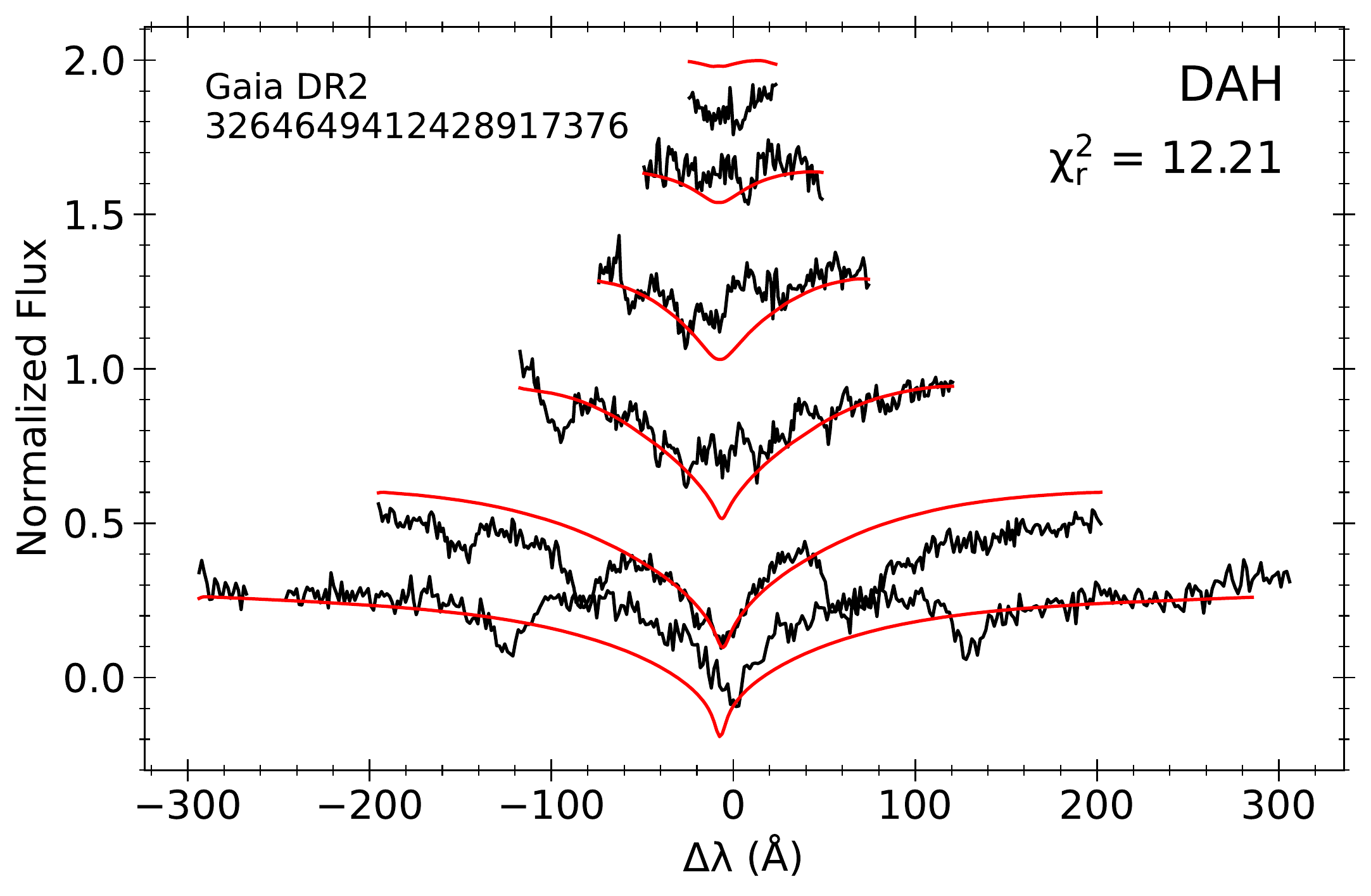}
    \includegraphics[width=\columnwidth]{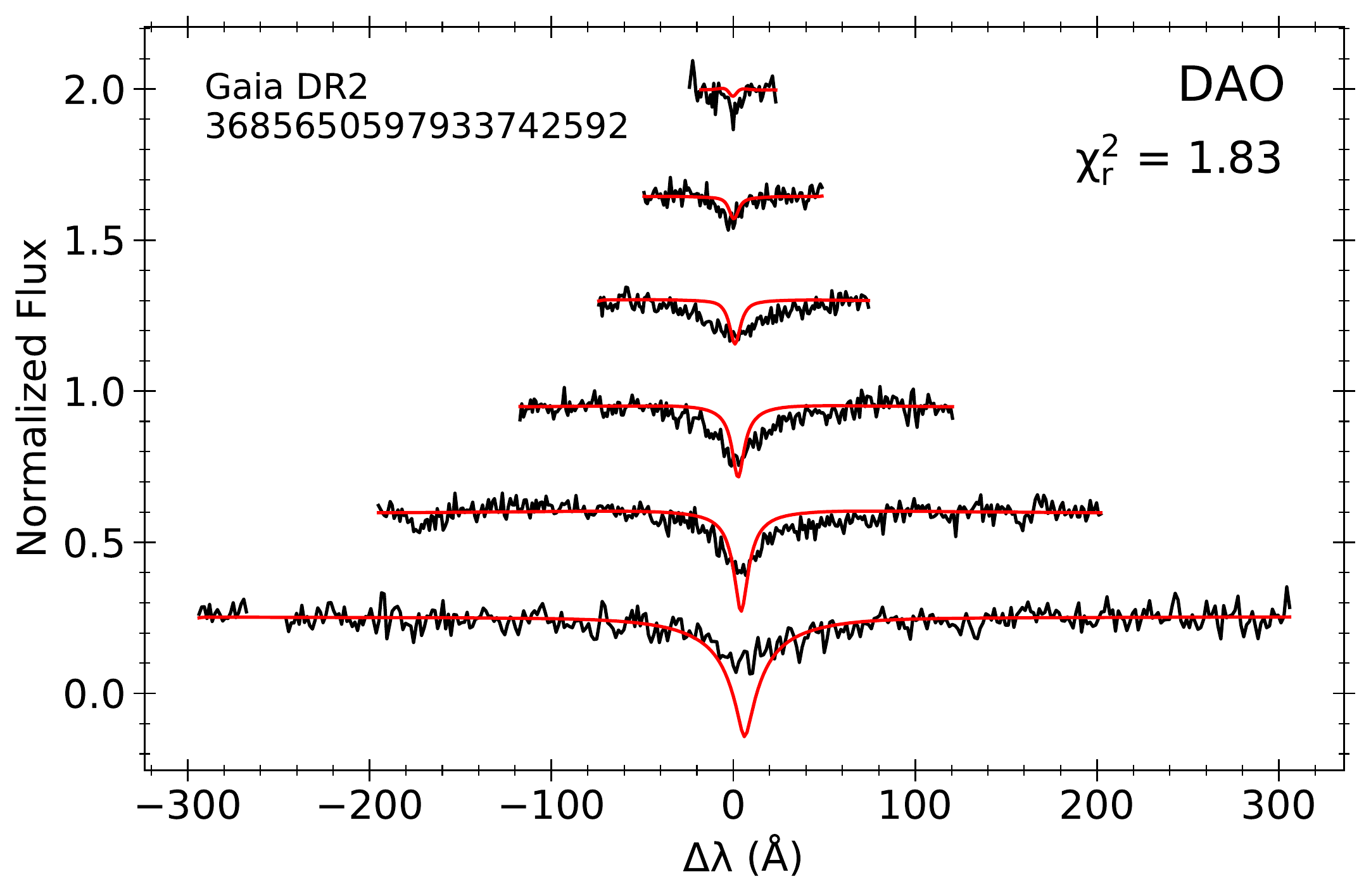}
    \caption{Example output of our generative fitting pipeline for a pure-DA spectrum (top) as well as two outlier white dwarfs (middle and bottom) from the SDSS catalog \citep{Kepler2019}. Balmer lines are arranged from H$\alpha$-H$8$ from bottom to top. The DAH is a magnetic white dwarf with symmetric Zeeman splitting of the Balmer absorption lines. DAO spectra have \ion{He}{ii} absorption lines and characteristically flattened Balmer lines. The latter spectrum shows very weak Helium absorption lines upon careful inspection, but was initially identified by the fact that our fitted spectroscopic stellar labels disagree significantly with its photometric labels \citep{Dufour2017}.}
    \label{fig:outliers}
\end{figure}

There are several ways to identify non-DA spectra that are more specific than the $\chi_r^2$ method, for example searching for particular absorption lines. Calcium lines are strong indicators of metal-polluted white dwarfs \citep{Farihi2010}, with the \ion{Ca}{i} resonance line being especially prominent in cool white dwarf spectra \citep{Blouin2019}. The traditional way to identify magnetized white dwarfs is visual inspection of the Balmer lines to search for Zeeman splitting, although this is only resolved for fields $\gtrsim 2$ MG on SDSS-resolution spectra \citep{Kepler2013}. Additionally, contaminant A/F type main sequence stars can be removed from samples of white dwarf spectra on the basis of their narrower Balmer lines \citep{Kepler2019}. However, methods that rely on visual inspection are outright infeasible for larger datasets such as those from upcoming spectroscopic surveys. Automating the detection of potentially interesting and outlier white dwarf spectra is the subject of ongoing research in our group.

Fitting the spectrum with a pure-DA model and inspecting the quality of fit and returned stellar labels from different observables can also can be useful to identify outlier spectra \citep{GB2019}. Significant disagreement between spectroscopic and photometric labels can be used to identify double-degenerates like DA+DA and DA+DC binaries. We have also verified, using simulated spectra of binaries, that increasing the Doppler shift between the two stars in a pair results in a gradual increase in the reported $\chi^2_r$ as expected. Our tool can be used in conjunction with the `WD-models' package \citep{wd_models} in Python to compare spectroscopic and photometric labels. Future iterations of our tool will include improved integration with `WD-models', facilitating the use of photometric priors for spectroscopic fits. Single lined DA+DA binaries can look exactly like single DA spectra, and therefore using a photometric label constraint is powerful for finding inconsistent spectra. We show some examples of outlier SDSS spectra we found using our tool in Figure \ref{fig:outliers}. \cite{GB2019} provide a comprehensive summary of how to use spectroscopic fitting techniques to find interesting white dwarf systems. 

As a real-world example of these applications, this year we obtained spectra for 5 high-mass DA candidates with the Double Imaging Spectrometer (DIS) at the Apache Point Observatory (APO). We based our selection on SDSS and Gaia data, looking for unusual high-mass white dwarfs with high velocities and/or photometric variability. We ran our generative fitting pipeline on these spectra and derived effective surface temperatures and surface gravities. Our labels for 2 targets were inconsistent with photometry. One of these (SDSS J105449.74+552307.9) was revealed by follow-up Gemini spectroscopy (program GN-2018B-FT-115, PI: Hwang) to a be a weakly magnetic white dwarf (DAH). Another candidate other exhibits interesting radial velocity variations that warrant further follow-up observations. This is an example of ways we have used our tool to find interesting white dwarf candidates.

\section{Discussion}\label{discussion}

In this paper we have developed and summarized two methods to derive stellar labels $T_{\text{eff}}$ and $\log{g}$ from spectroscopic observations of hydrogen-rich white dwarfs. Each has its advantages and disadvantages for different applications. The generative fitting pipeline (Section \ref{synthetic}) is a similar methodology to traditional fitting with synthetic models, albeit made much faster by the use of neural networks to generate synthetic spectra from a grid of pre-computed model spectra. This approach is best suited for single observations where an accurate characterization of the label posterior distributions is desired. This is certainly the most interpretable technique since it provides a visual goodness-of-fit to model spectra, $\chi^2$ statistics, and statistically precise labels and uncertainties consistent with past studies \citep{Tremblay2019}. 

Whilst the generative fitting pipeline is comparatively fast for single stars, it can require significant computational resources to fit a large dataset of stars. The parametric random forest (Section \ref{rf}) is a much faster method suitable for transfer learning, to derive labels for a large dataset when accurate labels are known for some subset of the dataset. However, some information may be lost when reducing the dimensionality from the spectrum to fitted line parameters. 

One crucial limitation of the methods described here is that they rely on the accuracy of the underlying theoretical models. The generative fitting pipeline is limited by the systematic uncertainties in theoretical synthetic spectra, and the random forests are additionally limited by any systematics in the training labels derived by other groups. Current theoretical limitations likely imply that spectroscopic $\log{g}$ cannot be constrained better than within 0.1 dex. An additional unresolved uncertainty is the `$\log{g}$ upturn' that causes low-temperature stars ($T_{\text{eff}}$) to have systematically overestimated $\log{g}$ measurements.

Our software tool is not intended to replace careful and intensive spectroscopic analysis with full-featured atmospheric models. Many scientific cases require more advanced models, for example those studying metal polluted or very cool white dwarfs \citep{Blouin2020}. However the proprietary and restricted nature of nearly all current white dwarf models necessitated a tool like ours to derive white dwarf atmospheric parameters. We will expand our tool's functionality as more white dwarf models become publicly available in the future. Our generative fitting pipeline can be trained on new model spectra that span a wider label space, and can be generalized to additional labels like magnetic field and metal abundances. Additionally, our random forest tool can be re-trained when larger spectroscopic datasets are fit by groups with proprietary model atmospheres.

In this work, we have summarized the methodology and accuracy of {\sc wdtools} and summarized several possible applications. We utilized {\sc wdtools} in our study of white dwarf gravitational redshifts \citep{Chandra2020}, using the generative fitting pipeline of Section \ref{synthetic} to derive $\log{g}$ for over three thousand DA white dwarfs. When combined with gravitational redshifts, this enabled us to empirically measure the mass--radius relationship of white dwarfs across a wide range of masses, the first measurement of its kind for such a large sample of stars. We hope the open-source availability of our tool, as well as future public releases of white dwarf atmospheric models, will enable more statistical studies of the ever-growing sample of known white dwarfs in our Galaxy. 

\section*{Acknowledgments}
	
 We thank the referee, Pier-Emmanuel Tremblay, for constructive comments and suggestions that significantly improved our manuscript. VC thanks Yuan-Sen Ting, JJ Hermes, Kirsten Hall, and Sihao Cheng for productive conversations and feedback on the analysis. We thank Detlev Koester for providing us with a grid of DB synthetic spectra to validate our software. VC is supported by the JHU PURA and DURA. VC and HCH are supported by Space @ Hopkins. VC, HCH, and NLZ were supported in part by NASA-ADAP 80NSSC19K0581.
 
Based in part on observations from the Sloan Digital Sky Survey. Funding for the Sloan Digital Sky Survey IV \citep{SDSS2017} has been provided by the Alfred P. Sloan Foundation, the U.S. Department of Energy Office of Science, and the Participating Institutions.
This work has made use of data from the European Space Agency (ESA) mission {\it Gaia} \citep{Gaia2016,Gaia2018}, processed by the {\it Gaia} Data Processing and Analysis Consortium (DPAC). Part of this research project was conducted using computational resources at the Blue Crab cluster of the Maryland Advanced Research Computing Center (MARCC). 

Based in part on observations collected at the European Southern Observatory, as well as on observations obtained with the Apache Point Observatory 3.5-meter telescope, which is owned and operated by the Astrophysical Research Consortium. Based on observations obtained at the international Gemini Observatory, a program of NSF’s NOIRLab, which is managed by the Association of Universities for Research in Astronomy (AURA) under a cooperative agreement with the National Science Foundation. on behalf of the Gemini Observatory partnership: the National Science Foundation (United States), National Research Council (Canada), Agencia Nacional de Investigaci\'{o}n y Desarrollo (Chile), Ministerio de Ciencia, Tecnolog\'{i}a e Innovaci\'{o}n (Argentina), Minist\'{e}rio da Ci\^{e}ncia, Tecnologia, Inova\c{c}\~{o}es e Comunica\c{c}\~{o}es (Brazil), and Korea Astronomy and Space Science Institute (Republic of Korea). 
This research has made use of the Spanish Virtual Observatory supported from the Spanish MINECO/FEDER through grant AyA2017-84089. 

\subsection*{Data Availability}

The Gaia, SDSS, and ESO UVES data used in this work to validate our software are freely available from their respective data archives (\url{https://gea.esac.esa.int/archive/}, \url{http://skyserver.sdss.org/dr16/en/home.aspx} and \url{http://archive.eso.org/wdb/wdb/adp/phase3_spectral/form} respectively). The Gemini spectra we observed during program GN-2018B-FT-115 are now publicly available from their archive (\url{https://archive.gemini.edu/searchform}). The software tool we present here is open-source and hosted on GitHub (footnote \ref{gh}) and Zenodo (\url{http://doi.org/10.5281/zenodo.3828686}). The DA synthetic spectra we use to train our model are hosted online (footnote \ref{SVO}), and the DB synthetic spectra provided to us by \cite{KoesterPrivate} are currently under restricted access. The Python notebooks used to generate the figures in this paper are available from the corresponding author upon request. 

\bibliographystyle{mnras} 
\bibliography{bib.bib}
\label{lastpage}
\end{document}